\documentclass[acmsmall]{acmart}

\AtBeginDocument{%
  }

\let\oldtodo\todo
\usepackage{todonotes}

\let\todo\oldtodo 

\usepackage[]{collab}
\usepackage{enumitem}
\usepackage{balance}
\usepackage{xcolor}
\usepackage{subcaption}
\usepackage{multirow}
\usepackage[most]{tcolorbox}
\usepackage{tabularx}
\usepackage{siunitx}
\usepackage{float}
\usepackage{hyperref}
\usepackage{caption}
\usepackage[commandnameprefix=always]{changes}

\def\name{\textsc{SCLogger}\xspace}

\definecolor{darkgreen}{rgb}{0.0, 0.5, 0.0}

\newcommand{\ie}{{\em i.e.},\xspace}
\newcommand{\eg}{{\em e.g.},\xspace}
\newcommand{\fixedwidth}[1]{{\ttfamily \small #1}}

\definecolor{ballblue}{rgb}{0.13, 0.67, 0.8}
\collabAuthor{yc}{orange}{Yichen Li}
\collabAuthor{yt}{blue}{Yintong Huo}
\collabAuthor{ry}{pink}{Renyi Zhong}
\collabAuthor{zh}{purple}{Zhihan Jiang}
\collabAuthor{jy}{ballblue}{Jinyang Liu}
\collabAuthor{jj}{cyan}{Junjie Huang}
\collabAuthor{jz}{magenta}{Jiazhen Gu}
\collabAuthor{pj}{green}{Pinjia He}

\ccsdesc[500]{Software and its engineering~Maintaining software}

\keywords{code generation, software maintenance, large language models}

\begin{document}

\title{Go Static: Contextualized Logging Statement Generation}

\author{Yichen Li}
\orcid{0009-0009-8370-644X}
\affiliation{%
  \institution{The Chinese University of Hong Kong}
  \city{Hong Kong}
  \country{China}
}
\email{ycli21@cse.cuhk.edu.hk}

\author{Yintong Huo}
\orcid{0009-0006-8798-5667}
\affiliation{%
  \institution{The Chinese University of Hong Kong}
  \city{Hong Kong}
  \country{China}
}
\email{ythuo@cse.cuhk.edu.hk}

\author{Renyi Zhong}
\orcid{0000-0001-6626-4437}
\affiliation{%
  \institution{The Chinese University of Hong Kong}
  \city{Hong Kong}
  \country{China}
}
\email{ryzhong22@cse.cuhk.edu.hk}

\author{Zhihan Jiang}
\orcid{0009-0003-1988-6219}
\affiliation{%
  \institution{The Chinese University of Hong Kong}
  \city{Hong Kong}
  \country{China}
}
\email{zhjiang22@cse.cuhk.edu.hk}

\author{Jinyang Liu}
\orcid{0000-0003-0037-1912}
\affiliation{%
  \institution{The Chinese University of Hong Kong}
  \city{Hong Kong}
  \country{China}
}
\email{jyliu@cse.cuhk.edu.hk}

\author{Junjie Huang}
\orcid{0009-0004-6962-5292}
\affiliation{%
  \institution{The Chinese University of Hong Kong}
  \city{Hong Kong}
  \country{China}
}
\email{junjayhuang@outlook.com}

\author{Jiazhen Gu}
\orcid{0000-0002-5831-9474}
\affiliation{%
  \institution{The Chinese University of Hong Kong}
  \city{Hong Kong}
  \country{China}
}
\email{jiazhengu@cuhk.edu.hk}

\author{Pinjia He}
\orcid{0000-0003-3377-8129}
\affiliation{%
  \institution{The Chinese University of Hong Kong}
  \city{Shenzhen}
  \country{China}
}
\email{hepinjia@cuhk.edu.cn}

\author{Michael R. Lyu}
\orcid{0000-0002-3666-5798}
\affiliation{%
  \institution{The Chinese University of Hong Kong}
  \city{Hong Kong}
  \country{China}
}
\email{lyu@cse.cuhk.edu.hk}

\renewcommand{\shortauthors}{Li et al.}
\begin{abstract}



Logging practices have been extensively investigated to assist developers in writing appropriate logging statements for documenting software behaviors. 
Although numerous automatic logging approaches have been proposed, their performance remains unsatisfactory due to the constraint of the single-method input, without informative programming context outside the method. Specifically, we identify three inherent limitations with single-method context: limited static scope of logging statements, inconsistent logging styles, and missing type information of logging variables.

To tackle these limitations, we propose \name, the first contextualized logging statement generation approach with inter-method static contexts.
First, \name extracts inter-method contexts with static analysis to construct the \textit{contextualized prompt} for language models to generate a tentative logging statement. 
The contextualized prompt consists of an extended static scope and sampled similar methods, ordered by the chain-of-thought (COT) strategy. 
Second, \name refines the access of logging variables by formulating a new \textit{refinement prompt} for language models, which incorporates detailed type information of variables in the tentative logging statement.

The evaluation results show that \name surpasses the state-of-the-art approach by 8.7\% in logging position accuracy, 32.1\% in level accuracy, 19.6\% in variable precision, and 138.4\% in text BLEU-4 score. Furthermore, \name consistently boosts the performance of logging statement generation across a range of large language models, thereby showcasing the generalizability of this approach.

\end{abstract}
\titlenote{Yintong Huo is the corresponding author.}

\maketitle

\section{Introduction}

Logging practices have been widely studied since logs provide rich resources for software debugging and maintenance~\cite{yuan2012improving,du2017deeplog,liu2023scalable,zhang2017pensieve}. A logging statement is typically comprised of three components~\cite{jiang2023large,jiang2023llmparser}: logging level, logging variables and logging text. The following shows an example of logging statement, where the terms \textit{``info''}, \textit{``service''}, and \textit{``Entry to state for''} represent the logging level, logging variable, and logging text, respectively. 
Furthermore, the logging variable \textit{``service''} is further utilized through the invocations of its member functions \textit{getServiceState()} and \textit{getName()}. 

\begin{center}
\small
    \fbox
    {\shortstack[l]{
    \fixedwidth{LOG.info("Entry to state " + service.getServiceState() + " for " + service.getName());}
    }}
\end{center}

To facilitate developers writing logging statements, a number of works are proposed to build models for automated logging statement generation. These works focus on two aspects of logging, including generating the logging contents (i.e., \textit{what-to-log}) and suggesting the logging positions (i.e., \textit{where-to-log}) in the code context. 
A preliminary work ~\cite{he2018characterizing} pointed out that logging statements in one project usually share similar patterns so that the logging history can be learned to generate new logging statements. Motivated by such patterns, existing logging generation approaches are mostly neural-based and supervised, learning logging patterns from historical data. These approaches can be further categorized into two types: discriminative and generative.

The idea of the discriminative logging approach is to develop deep-learning models for determining \textit{single} component in a logging statement or predicting \textit{whether} a logging statement should be contained in a piece of code. 
For example, DeepLV~\cite{liu2022tell} suggests log levels by building a Bi-LSTM network to learn syntactic code features and log message features. 

However, these discriminative logging models are restrained by their classification nature, which relies on a pre-defined set of classes and cannot generate complete logging statements.
Inspired by trending language models, recent generative logging models overcome these limitations by considering the task of logging statement generation as the problem of text generation, which accepts code snippets and outputs the entire logging statement and its corresponding logging place. The pioneer logging statement generative approach, LANCE~\cite{mastropaolo2022using}, employs a Text-to-Text Transfer Transformer (T5) mode~\cite{raffel2020exploring} to inject complete logging statements given a code snippet. The modern large language models (e.g., GPT-3.5) also show promising results in this task~\cite{li2023exploring}.

While generative models are investigated for better performance, their analysis scopes for logging statement generation are still outstretched: \textit{When recommending logging statements for a specific method, existing approaches solely look into this method (or even a code block inside) while ignoring the programming context from other methods, not to mention from other files.} In particular, these models simply choose the method-level context (i.e., single-method) while ignoring the critical context outside the target method for inferring the logging position and corresponding logging statement. In the following, we present three significant inherent limitations of the previous single-method context and further discuss them with real-world examples in Sec.~\ref{sec:study}.

(1) \textit{Limited static scope of logging statements}.
Complete software integrates numerous interconnected methods, each of which serves as a function being called by others. The flow of execution across various methods offers a comprehensive overview of code functionalities in the entire system for developers. Relying on a limited single-method context, it is challenging to infer the logging purpose and thus decrease the logging quality. Apart from execution flow, the available variables that are beyond the scope of the target method (e.g., attributes in the current class) are also indispensable. Without knowing the available variables of given method-level context, choosing logging variables beyond the method scope becomes nearly impossible.


(2) \textit{Inconsistent logging styles}. 
In well-maintained software projects, consistent logging styles are crucial~\cite{rong2018logging,rong2020can,rong2017systematic,yuan2012characterizing} for ensuring log readability~\cite{li2023they} and the coherence of logs. 
Such consistency encompasses maintaining coherent logging levels for component lifecycles~\cite{liu2022tell}, choosing appropriate words~\cite{li2023they}, and using accordant log text separators.
Previous works~\cite{ding2022logentext,ding2023logentext} have shown similar code can provide additional information on the syntactic structure of logging text and logging pattern. 
However, in the method-level context, learning the logging style of a specific project becomes challenging without in-project adaptation. 
Relying solely on the general knowledge that pre-trained models acquired during the large-scale pre-training phase poses a challenge in providing project-specific, consistent logging text and appropriate logging levels for a given logging statement. 
This could lead to inconsistencies in the logging style, potentially impacting the readability~\cite{li2023they} and maintainability of the software.

(3) \textit{Missing type information of logging variables}.
Suggesting proper logging variables not only requires predicting the object variables (i.e., \textit{service} in the first paragraph) themselves but also predicting their attributes and member functions (i.e., the \textit{getServiceState()} member function). However, variable attributes and member function declaration are usually defined within the class, which stays out of the target method context. 
Existing work~\cite{mastropaolo2022using, liu2019variables}
only focus on the inside content of a method while never covering the detailed variable type information from the outside. This type information, if present, could provide explicit definitions for the attributes and member functions of the variables. 
Without such information, the logging models may mistakenly invoke non-existent member functions and misuse variables, which will further lead to compilation errors and software bugs.


\textbf{Our Work.} 
To tackle these limitations, we propose \name, the first logging statement generation approach powered by inter-method \textbf{\underline{S}}tatic \textbf{\underline{C}}ontexts.
\name~analyzes inter-method programming contexts for logging statement generation with four phases, including static scope extension, logging style adaption, contextualized prompt construction, and logging variable refinement. 
In the static scope extension phase, \name~extends the static scope of the given method by constructing the function invocation relationships, deriving the execution paths containing logging statements, as well as collecting available variables for the current method, such as class member variables and inherited variables, as the logging variable candidates. During this phase, the limitation of \textbf{limited static scope} is mitigated. 
In the logging style adaption phase, \name~adapts the idea of the in-context learning (ICL) strategy to select intra-project similar examples demonstrating the logging styles and logging patterns of the current project, which address the issue of \textbf{inconsistent logging style}.
In the contextualized prompt construction phase, \name~translates the inference steps of logging statement generation into a chain-of-thought (COT) prompt. The context-aware code knowledge generated by the previous two steps, along with the COT prompt, is put together as a \textit{contextualized prompt}. With the constructed contextualized prompt, \name~then invokes a large language model to generate the required logging statement for the given target method. 
In the final phase, i.e., logging variable refinement, \name further refines the usage of logging variables. It provides comprehensive type definitions for logging variables that were generated in the third phase. This allows \name to self-refine and correct the variable usage, ensuring syntactical correctness of the generated logging statement. This phase tackles the limitation of \textbf{missing type information}.

Following the previous works~\cite{ding2022logentext,ding2023logentext}, we conduct a comprehensive evaluation on ten open-source Java projects from different domains. 
The results show that \name achieves the best performance over all metrics in both deciding logging location (i.e., \textit{where-to-log}) and generating logging statement content (i.e., \textit{what-to-log}). More specifically, \name outperforms the state-of-the-art approach by 8.7\% in logging location accuracy, 32.1\% in logging level accuracy, 19.6\% in logging variable precision, and 138.4\% in logging text \textit{BLEU-4} score, respectively.

Moreover, \name consistently enhances the performance of logging statement generation models with various backbone large language models, thus demonstrating its generalizability. Besides, we explore the individual contribution of each phase and provide the reasoning for the improvements brought by our approach.

This paper’s contributions are summarized as follows:
\begin{itemize}[leftmargin=*, topsep=0pt]

\item To the best of our knowledge, we propose the first contextualized logging statement generation approach named \name. With analyzed static context, \name addresses the limitations of current method-level approaches with limited context.

\item We propose a novel prompt structure to incorporate static context of code into large language models, which can be generalized to various language models for future improvement.

\item We conduct the comprehensive evaluation of \name on public logging datasets. The results demonstrate the effectiveness of \name and the adaptability of \name with different backbone models.
\item The source code of \name is publicly available at \href{https://github.com/YichenLi00/SCLogger}{https://github.com/YichenLi00/SCLogger} to benefit both developers and researchers.
\end{itemize}

\section{Motivating Study}\label{sec:study}

Logging statement automation has been a longstanding area in the domain of software development and maintenance~\cite{he2021survey}, since high-quality logging statements can precisely describe system activities and ease the burden for maintainers to diagnose system anomaly behaviors.
A wide range of approaches have been proposed to automatically recommend proper logging points~\cite{mastropaolo2022using,li2020shall,zhao2017log20} or generate effective logging statements ~\cite{mastropaolo2022using,he2018characterizing,ding2022logentext,liu2022tell,ding2023logentext} for developers.

However, we discover that existing studies contain the same limitation of only exploiting single-method information for automatic logging, missing the inter-method contexts. In fact, since methods in software are interrelated, the programming context across different methods plays a vital role in understanding logging purposes and making logging suggestions accordingly.
Intuitively, if the variable is defined outside a certain method, the developer can only log this variable properly once he/she reads the outside context.
In this section, we perform a motivating study to illuminate the need for a more context-aware approach. In particular, we present real-world cases to illustrate limitations introduced by using a single-method context, that is, \textit{limited static scope}, \textit{inconsistent logging style}, and \textit{missing type information of variables.}

\subsection{Limited static scope}\label{sec:example1}

\begin{figure*}[tbp]
    \centering
    \includegraphics[width=\textwidth]{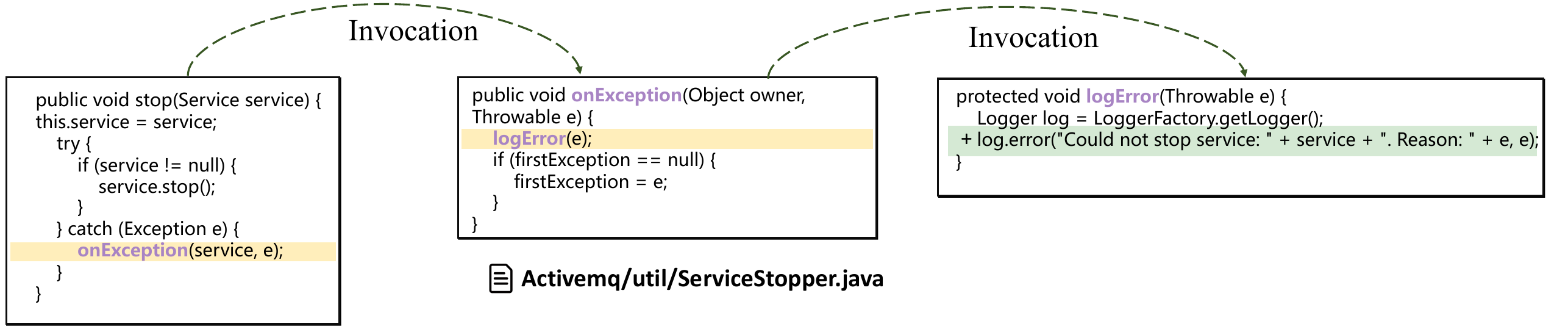}
    \caption{Motivating example 1. The origin logging statement is highlighted in the \textcolor{darkgreen}{green} area while the invocation points are highlighted in the \textcolor{orange}{orange} area.} 
    \label{fig: example1}
    \vspace{-12pt}
\end{figure*}

Modern software consists of a great number of methods, each of which is responsible for small functionalities. Single methods are often short, and their content is insufficiently informative for logging generation models to grasp the logging purpose. The static profile from other methods, including invocations, subsequent logs in execution order, and available variables, should be expanded within the current static scope. This additional context can be instrumental in understanding system behavior for logging purposes.

We exhibit Fig.~\ref{fig: example1} as an example from the ActiveMQ project to illustrate the significance of static scope. In this example, the \texttt{logError} method, which consists of two lines, is expected to log the error information for a certain error. Given only the method name and a single line of logger registration, it becomes relatively challenging to predict what kind of error should be logged~\cite{yuan2012improving}. However, by tracing the method's invocation sequence, we discover that \fixedwidth{logError} is called within the method \fixedwidth{OnException}, which in turn is invoked by the \fixedwidth{stop} method of the \fixedwidth{ServiceStopper} class. Consequently, such invocations reveal that the purpose of \fixedwidth{logError} method is to record the state of service stop from \fixedwidth{stop} method. Without the invocation information of the \fixedwidth{logError} method, it would be nearly impossible to infer the appropriate logging statements for recording such service \fixedwidth{stop} behavior and record this error. Moreover, the fact that \textit{service} is also beyond the scope of the \textit{logError} method further underscores the limitations of method-level static scope.

Furthermore, many invocation relationships span across different files, and some logs need to be interpreted within a sequence of logs~\cite{du2017deeplog,li2022intelligent,chen2021experience,huo2023autolog}.
Therefore, directly extending context to the file level may also be insufficient to address the issue of lack of information.
To thoroughly understand the relative functionality of target method in program execution, it is essential to acquire a more comprehensive and specific context through inter-procedural analysis.
\vspace{-10pt}

\begin{figure*}[htbp]
    \centering
    \includegraphics[width=\textwidth]{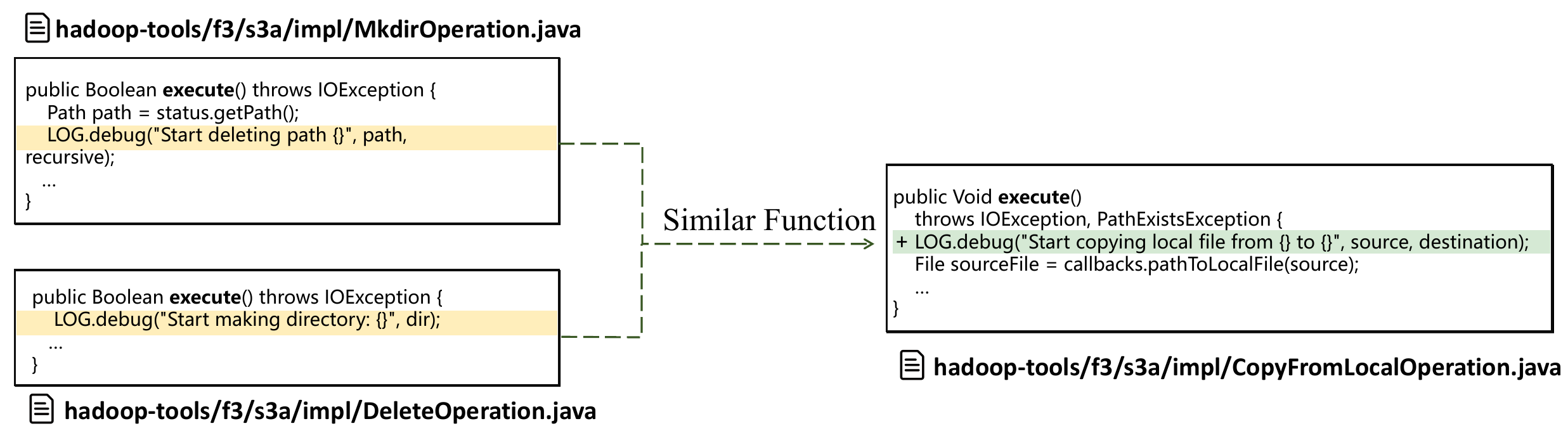}
    \caption{Motivating example 2. The origin logging statement is highlighted in the \textcolor{darkgreen}{green} area while the logging statements in the similar methods are highlighted in the \textcolor{orange}{orange} area.} 
    \label{fig: example2}
    \vspace{-5pt}
\end{figure*}

\subsection{Inconsistent logging style}\label{sec:example2} Logging style in software development and maintenance is maintained with relative consistency~\cite{rong2018logging,rong2020can,rong2017systematic,yuan2012characterizing} in a mature project. Examples of this consistency include maintaining coherent logging levels for the lifecycle of certain components~\cite {liu2022tell}, writing logging statements with similar words~\cite{li2023they}, applying the same separators in logging text, and so on. To understand the logging style of the current project, developers naturally learn from similar logging examples within the project.

Fig. ~\ref{fig: example2} showcases an instance from the Hadoop-AWS toolset within the Hadoop project. The three \fixedwidth{execute} methods, extracted from \fixedwidth{MkdirOperation}, \fixedwidth{CopyFromLocalOperation}, and \fixedwidth{DeleteOperation} files, are the primary execution methods that enable various file operations in the S3A filesystem. These methods serve as similar methods, exhibiting similar logging styles, particularly in terms of levels and wording. Without similar methods as a reference of logging style, it would be challenging to infer that such file operations should choose the \fixedwidth{debug} level and use the \fixedwidth{Start doing} text structure. Furthermore, given only the target function \fixedwidth{execute}, the model has no way of knowing that similar file operation methods in the project will log \fixedwidth{Start doing..} at the beginning of the methods before performing file operations. Hence, missing additional methods as references might cause logging style inconsistency, which can be mitigated by providing samilar methods from the same project.


\begin{figure*}[htbp]
    \vspace{-5pt}
    \centering
    \includegraphics[width=\textwidth]{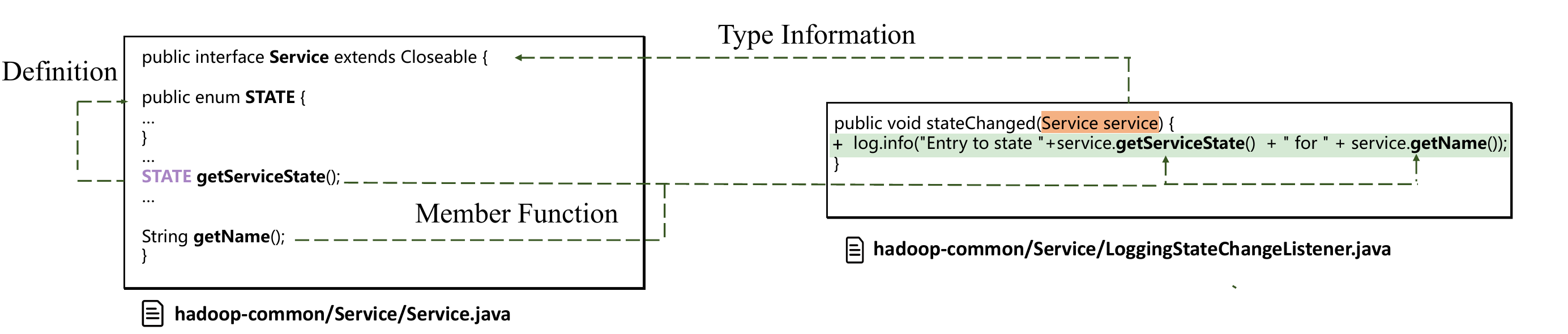}
     \vspace{-20pt}
    \caption{Motivating example 3. The origin logging statement is highlighted in the \textcolor{darkgreen}{green} area while the corresponding logging variable is highlighted in the \textcolor{orange}{orange} area. } 
    \label{fig: example3}
    \vspace{-10pt}
\end{figure*}

\subsection{Missing type information of logging variables.}
Since variables are often defined outside the method (\eg class attributes for object-oriented programming languages~\cite{yuan2012improving}), the third limitation of the intra-method context for logging is the missing type information of logging variables. Ignoring such information obstructs logging models from determining the proper usage of a variable (e.g., properties from a class) even though they understand the logging purpose. 

Fig.~\ref{fig: example3} showcases an example from the Hadoop project, emphasizing the critical role of variable type information context within the automated logging process. The primary objective of this logging statement is to record the state change status of the given service. To this end, it is essential to invoke the two member functions (i.e., \texttt{getServiceState()} and \texttt{getName()}) from other classes defined outside the method, which retrieve the defined service state and name, respectively. 
Without integrating the \texttt{Service} interface information at \texttt{Service.java} into logging context,
logging models are asked to guess the member functions of \texttt{service}, inevitably impairing their performance and practicality.
The incorrect variable predictions (e.g., devoid invocation) can further lead to program compilation errors and software bugs.



\begin{tcolorbox}[boxsep=1pt,left=2pt,right=2pt,top=3pt,bottom=2pt,width=\linewidth,colback=white!95!black,boxrule=1pt, colbacktitle=white!30!black,toptitle=2pt,bottomtitle=1pt,opacitybacktitle=0.4]
\textbf{Insights.} 
The motivating study demonstrates the limitations of method-level context in understanding the semantics of the target method, maintaining consistency in the project-specific logging style, and selecting appropriate logging variables. Hence, we should devise models equipped with more context-aware code knowledge that does not exist in the target method for effective logging.
\end{tcolorbox}




\begin{figure*}[tbp]
    \centering
    \includegraphics[width=\textwidth]{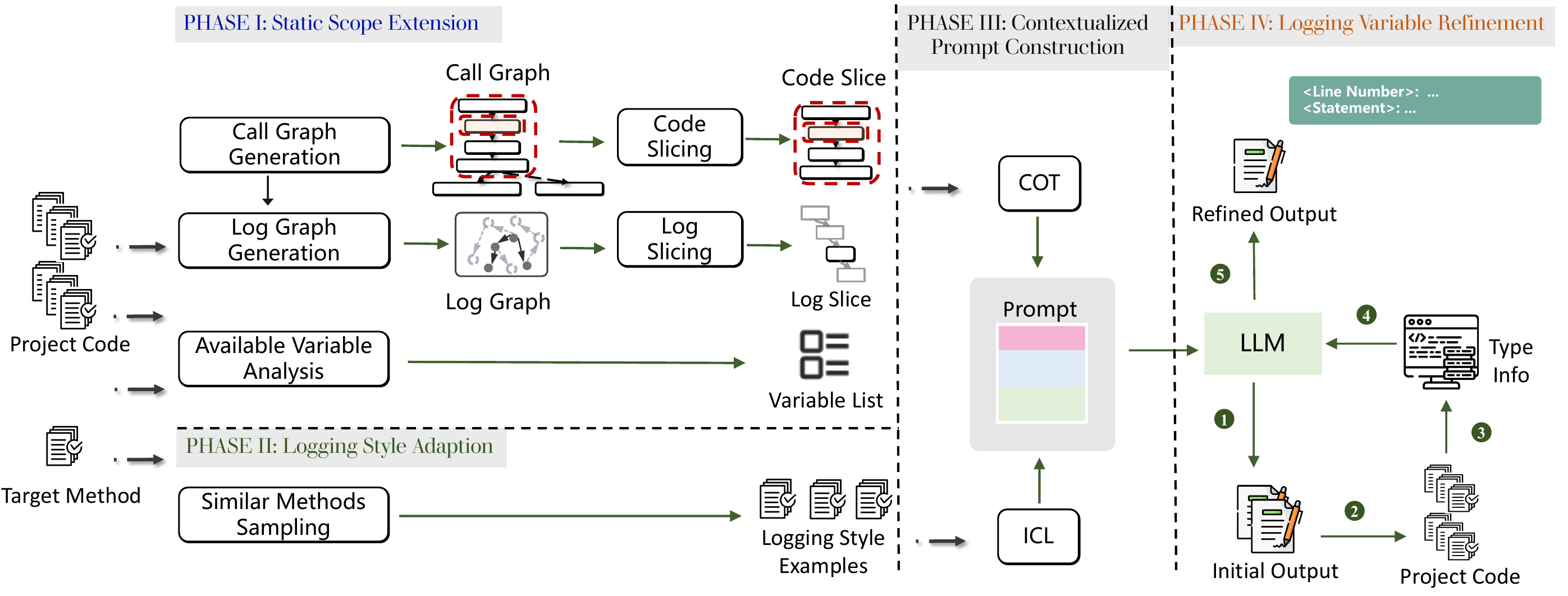}
    \vspace{-20pt}
    \caption{The overview workflow of \name.} 
    \label{fig: overview}
    \vspace{-20pt}
\end{figure*}

\section{Methodology}
\subsection{Overview}
First of all, we describe the problem of logging statement generation as follows.
Given a method code as input (i.e., target method), the goal of this task is to predict the logging location and its corresponding logging statement content. Specifically, the purpose of \name~is to predict the code line number (i.e., location) and generate a complete logging statement accordingly.



We propose \name, a static analysis enhanced contextualized logging statement generation framework via large language models. 
Intuitively, simply putting the entire project code into language models should work, however, such long input can be out of the model's input limit size and make the model get lost. As a result, we need a more advanced approach to extract useful information from the entire project.
To this end, \name extracts the logging-related context surrounding the target method and constructs contextualized prompts, which will be fed into language models for predicting logging position and generating logging statements. 
We present Fig.~\ref{fig: overview} to illustrate the workflow of \name.

\name takes the target method and its corresponding project code as input. The static scope extension phase derives inter-method information including code slice, log slice, and variable list. Code slice is a chain of methods code reflecting the method calling relationship associated with target methods. Log slice recording the potential subsequent and precedent logs during execution. The variable list contains all available variables for the target method. Afterwards, the logging style adaption phase utilizes the in-context learning (ICL) strategy by sampling a small set of similar methods from the project as logging style examples. Then, the third phase applies the chain-of-thought (COT) strategy~\cite{wei2022chain} to translate logging inference into a few steps, then combines it with the context-aware knowledge coming from the two previous phases, to form a \textit{contextualized prompt}. This combined prompt is then fed to the large language model (LLM) to get a tentative logging statement with the corresponding position (i.e., line number). During the final phase, logging variable refinement, \name constructs a new \textit{refinement prompt} that contains the detailed type information of the variable extracted from the tentative result. \name eventually feeds the new prompt into the LLM and generates the final logging statement with rectified variables.
The example with the contextualized prompt and the refinement prompt is illustrated in Fig.~\ref{fig: prompt}.


\vspace{-7pt}
\subsection{Static Scope Extension}
The static scope extension phase aims to extract the static context associated with logging that surrounds the target method. Ultimately, this phase will generate three types of context: the code slice, the log slice, and the list of available variables.

\subsubsection{Code Slice Generation}
To enhance the model's understanding of the target method's relative position and functionality within the project, we designed the code-slicing step to extract the invocation context.
The code slicing phase of \name is designed to extract the relevant invocation methods of target method $m_t$ from the statically generated call graph, which describes the invocation relationships among methods. To construct a relatively accurate call graph, our model utilizes a context-sensitive pointer analysis~\cite{li2018precision} to increase the precision of the call graph, especially in handling virtual method calls and similar situations.

In particular, the code slicing process identifies the methods that either invoke the target method or are invoked by the target method within two hops. Here, a single invocation can be considered as one hop. The position in the call graph corresponds to the node linked with the target method $m_t$, as per the invocation details. The graph traversal process follows the graph's directed edges, either forward or backward, to extract the preceding and succeeding methods. If there are too many paths within two hops, we only randomly select one of these invocation paths. As shown in Fig.~\ref{fig: slice}, for method \textit{get}, its method invocations of within two-hop are ~\textit{configureJob}, ~\textit{submit} and ~\textit{submitJob}
The output of code slicing is text descriptions of invocation relationships between the identified methods, as well as the chain of these method codes (i.e., \textit{get},~\textit{configureJob},~\textit{submit}, and ~\textit{submitJob}).

\subsubsection{Log Slicing}\label{sec:log_graph}

A single log event, being part of the log sequence, cannot be fully understood in isolation. The log slicing phase aims at identifying preceding logs and subsequent logs of a given target method based on a log graph which indicates the log dependency relationships.

\begin{figure*}[tbp]
    \centering
    \includegraphics[width=\textwidth]{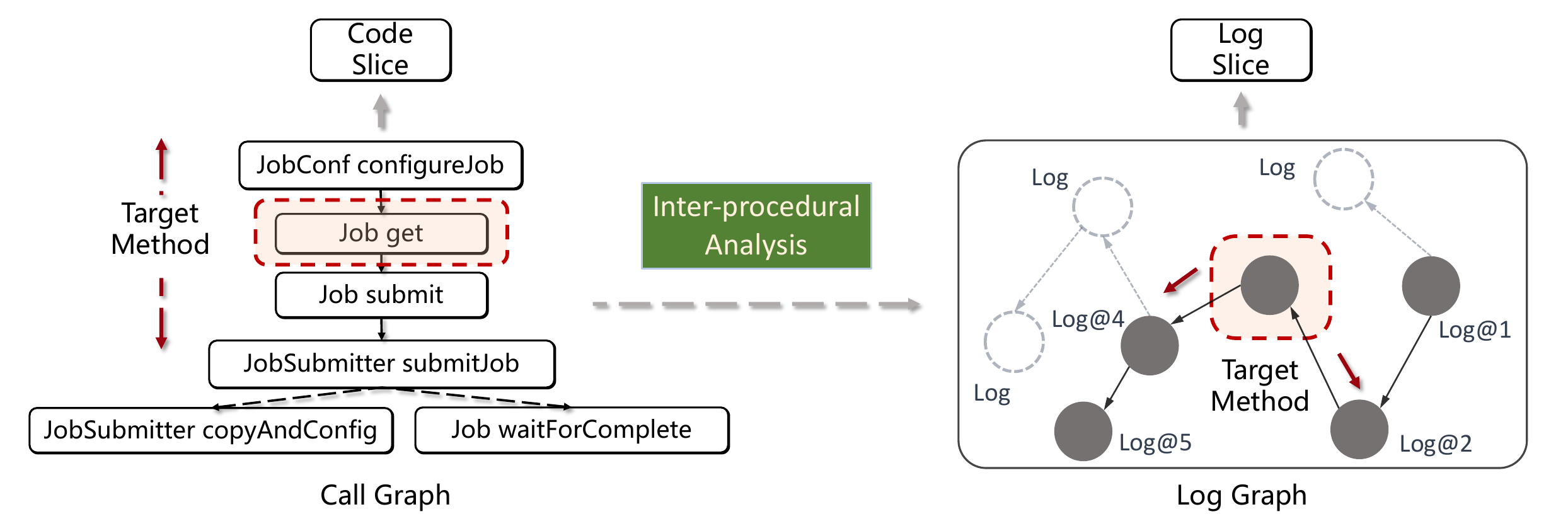}
    \vspace{-20pt}
    \caption{The log slice and code slice example of \name. The target method is highlighted in the \textcolor{red}{red} area.} 
    \vspace{-10pt}
    \label{fig: slice}
\end{figure*}

(1) \textit{Log Graph Construction}: To extract both preceding and succeeding logging statements for the target method, \name constructs log graphs for a given project $p$. This construction process is guided by the frameworks and ideas presented in previous works~\cite{huo2023autolog,chen2018automated,zhao2023studying}. A log graph is characterized as a directed graph $(L,E)$ (as shown in Fig.~\ref{fig: slice}), where $L$ represents the set of logging statements which is the node set in graph and $E$ represents the edge set, which is composed of program execution paths. Each logging statement $l \in L$ is obtained through static analysis of the source code of project $p$ with its belonging method $m$. Consequently, the connected logging statements in the log graph are causally related with the possible execution order, as they are derived from the execution paths within the same source code of project $p$.

Specifically, \name extends the framework of previous works~\cite{huo2023autolog,zhao2023studying}. It begins by analyzing and identifying the set of methods $M' \subseteq M$ that contain any logging statement. \name then uses the call graph for the project $p$ described above. For the call graph of project $p$, \name prunes the methods that do not directly or indirectly invoke any method $m \in M'$ from the complement of $\overline{M'}$ in $M$. These are referred to as log-irrelevant methods~\cite{huo2023autolog}. For the remaining methods, \name conducts the static execution path analysis~\cite{huo2023autolog} for each method. All relationships, including control flow and method calls, which are also part of the Interprocedural Control Flow Graph (ICFG), are integrated into the edge set $E$. Thus, each edge $e \in E$ represents a potential execution path from one logging statement to another, which aids in understanding and handling the dependency relationships among logging statements.

While the log graph provides a comprehensive set of potential executable paths, it includes certain paths that remain infeasible regardless of the constrains. To refine the log graph and reduce these infeasible paths, we undertake a preliminary intra-procedural constraint analysis. This process involves the collation of constraints in a method and the initial filtering out of any paths that contain unsatisfiable constraints. If a path within a method is determined to be infeasible, all paths reliant on this infeasible path are subsequently removed. Note that due to the inherent limitations, some potentially infeasible paths may still remain.

(2) \textit{Log Slice Generation}: With the log graph, \name generates a log slice that consists of preceding and succeeding logging statements for the target method $m_t$. This is accomplished by traversing the graph both backward and forward, beginning from the position of the target method in the log graph. As shown in Fig.~\ref{fig: slice},  log-specific slicing allows \name to capture long-distance log dependency information within two log hops (which will be beyond two method hops, as some methods may not contain logs) for the target method, within a greatly reduced and focused context. This is more targeted than directly incorporating all relevant code into the context according to the invocation sequence. The position in the log graph refers to the node associated with the target method $m_t$ according to invoke information and execution information. The process of traversing the graph involves following the directed edges in the graph, either forward or backward, to identify the preceding and succeeding logging statements. If an excessive number of nodes exist within two hops, we simply choose one of these paths from the methods available in the \textbf{training set}.

\subsubsection{Scope Variable Analysis}
The scope of accessible logging variables for a given method is not limited to its parameters and local variables. It also encompasses variables that are beyond the class level, including those inherited from a parent class. Consequently, merely capturing all member variables in the current file does not yield a comprehensive overview of available variables~\cite{yuan2012improving}. We illustrate the details of this phase as follows:

In the context of a target method $m_t$, we define several sets of variables. $V_p$ represents the set of parameters, which are the inputs to the method. $V_m$ stands for the set of local variables, which are defined and used only within the method. $V_c$ is the set of class member variables, which belong to the class that the method is a part of. $V_s$ is the set of static variables, which belong to the class as a whole rather than any specific instance. Lastly, $V_i$ denotes the set of inherited variables, which are class member variables that come from the parent class.

The logging process should focus on any variable $v$ that belongs to one or more of these sets ($v \in V_p \cup V_m \cup V_c \cup V_s \cup V_i$) during the execution of $m_t$~\cite{yuan2012improving}. As a result, all these variables need to be included in the context for selection as potential logging variables and subsequently form the available variable list.

Note that here, we are not giving the model the detailed type definition for each variable. The list of available variables primarily includes the variable name and roughly inferred type name. We will further discuss how we address the further type issue of logging variables in Sec.\ref{sec:refinement}.

\subsection{Logging Practice Adaptation}
To align with the logging style of the current project, \name employs the in-context learning (ICL) strategy, which has proved its effectiveness in code-related tasks~\cite{peng2023generative,gao2023constructing}, to adapt to the project's logging style. The in-context learning strategy~\cite{dong2022survey}, as its name suggests, allows the model to learn and adapt to the specific examples of the project, ensuring that the generated logging statements are in line with the project's existing style. Specifically, this strategy provides a few examples sampled from the inter-project training set (with labels, detailed in Sec.~\ref{sec:subjects}) as demonstrations of logging style so
that \name can learn from these examples to generate consistent logging statements.

Following previous works~\cite{peng2023generative,gao2023constructing}, we use the BM25~\cite{robertson2009probabilistic} similarity function to select these examples. The BM25 function is based on the TF-IDF (Term Frequency-Inverse Document Frequency) method. With the code of a target method as input, the BM25 function calculates the term frequency of each keyword in the query within the examples. It then multiplies this frequency by the inverse document frequency of the given term. The BM25 similarity score will be higher if there is a greater relevance between the query and the examples. This score helps \name to select the most relevant examples from the training set. Specifically, we select the top five examples from the training set with the highest BM25 similarity scores. The example prompts are  combined with the other prompt to form the complete contextualized prompt, as shown in Fig. ~\ref{fig: prompt}.

\subsection{Contextualized Prompt Construction}

\begin{figure*}[tbp]
    \centering
    \includegraphics[width=\textwidth]{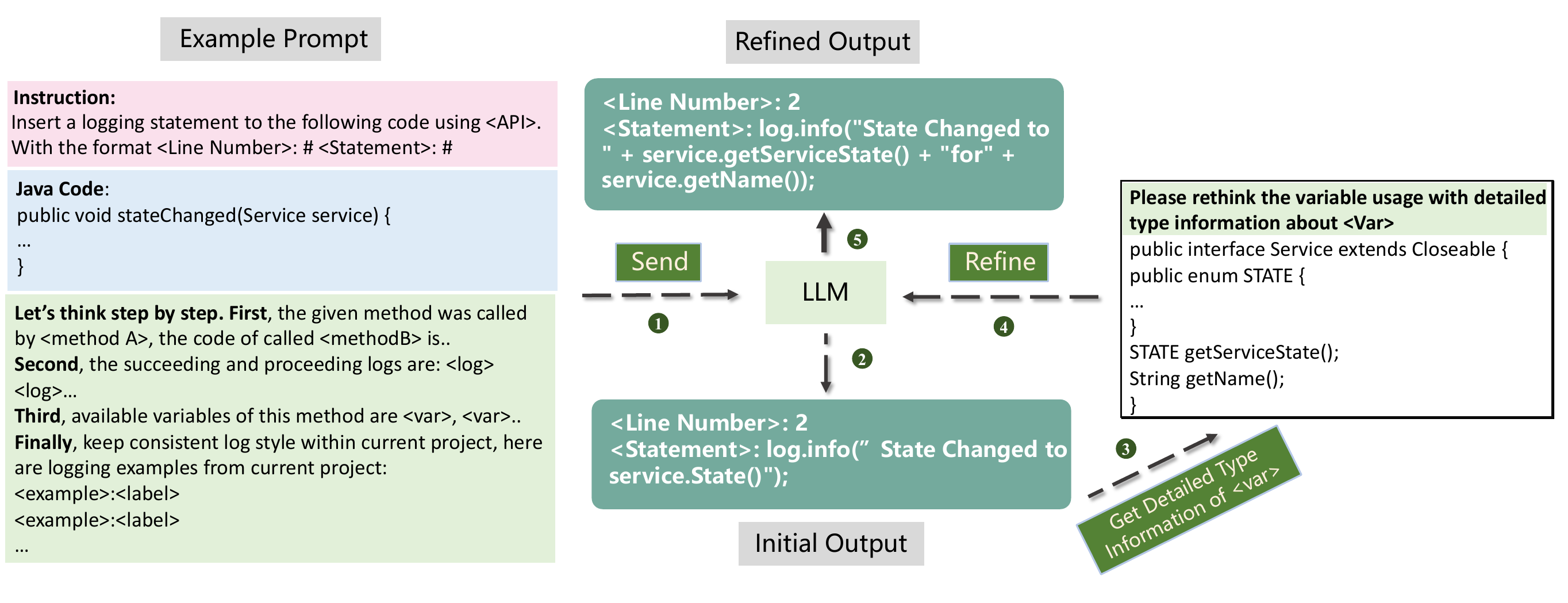}
    \vspace{-20pt}
    \caption{Contextualized prompt example with logging variable refinement.} 
    \label{fig: prompt}
    \vspace{-13pt}
\end{figure*}

As shown in Fig~\ref{fig: prompt}, \name converts all the context information gathered from static analysis during the first two phases into chain-of-thought (COT)~\cite{wei2022chain} prompts, which incorporate static domain knowledge regarding the requirements of a logging statement. The innovative COT approach, which employs sequential reasoning, guides language models towards generating complex and specific outputs~\cite{peng2023generative}. This approach allows the model to focus on one aspect of the task at the time, potentially enhancing the quality of the generated outputs.

Additionally, \name integrates the sampled logging style examples from the second phase with the In-Context Learning (ICL) strategy into the input prompt with reasoning prompt. It suggests the model to maintain a style consistent with the logging samples derived from the current project. By fusing these two components, the initial contextualized prompt is ultimately constructed.

\subsection{Logging Variable Refinement}\label{sec:refinement}

To address the issue of variable usage within selected variables, an intuitive idea might be to provide the detailed type information for every variable alongside the available variable list during the first phase. However, offering the detailed type definition of every available variable within the current scope is unrealistic and would lead the model into a wild-goose chase. 

To tackle this, we employed a two-staged variable type refinement mechanism to determine the proper usage of logging variables, as shown in Fig.~\ref{fig: prompt}. In the first stage, after providing the model with the contextualized prompt with an available variable list, we let the model conduct the inference. Then, we extract the logging variable chosen by the model and conduct a thorough type analysis of that variable to generate the detailed type information extracted from the project, which is then fed back to the model for reconsideration and self-refinement.

To obtain the detailed type information of a selected logging variable, we performed static analysis with class definition resolution. Given a variable $v$ and target method code $m_t$ within a project $p$, we first find the type $t$ of $v$ using with variable type resolving, then acquire its detailed information of $t$. Resolving the variable's type within method $m_t$ involves checking the parameter scope of $m_t$, local scope within $m_t$, and the class $c_t$ where $m_t$ is defined if the variable is a class-level variable. 

After obtaining type $t$, we denote the set of acquired referring class definitions in the project $p$ as $Def$ by analyzing imported intra-project packages and the current package. A class definition $def \in Def$ is a tuple $(t, M, A)$, where $t$ is the type, $M$ is the set of member functions, and $A$ is the set of attributes. The class definition resolving function $R: t \times p \rightarrow (M, A)$ is then defined as follows:

\begin{equation}
R(t, p) = (M, A) \quad \text{if} \quad \exists (t', M, A) \in Def \quad \text{such that} \quad t = t'
\end{equation}

The function $R$ takes a type $t$ and a project $p$ as input and returns the member functions $M$ and attributes $A$ of the type if there exists a definition $(t', M, A)$ in $Def$ where $t = t'$.

In the case of inheritance and polymorphism, we traverse the class hierarchy to collect all relevant member functions and attributes. If the class implements any interfaces, we also consider the methods declared in these interfaces. For generic classes, we consider all possible concrete types the generic type can take. This process is repeated recursively until we have a complete picture of the class's definition, including its inherited and overridden member functions and attributes.

In conclusion, we extract detailed type information of a variable, including its member functions and attributes, through variable type resolution and class definition resolution, for refining the generated logging statement.

The resolved detailed type information will be fed into \name, if the model realizes it has used the variable type incorrectly (e.g., not using \textit{var.getinfo()}), it will take this opportunity to carefully read the logging variable's information and correct the type error and further method usage. This allows \name to self-refine and
correct the variable usage, ensuring syntactical correctness of the generated logging statement.

\section{Experiment Setup}
\subsection{Subject Projects}\label{sec:subjects}
Following previous works~\cite{ding2022logentext,he2018characterizing,ding2023logentext}, we evaluate \name on ten open-source Java projects that span various domains, such as storage, cloud platforms, computation engines. Detailed information of these projects can be found in Table.~\ref{tab:subjects}. The source lines of code (SLOC) for the investigated projects range from 330K to 2.12M. Each project contains between 1,978 and 15,744 logging statements within 901 to 7,365 methods that contain these statements. Notably, every project boasts a development history exceeding ten years, which highlights the progression and evolution of each software system. Note that our decision not to select another generic logging dataset~\cite{mastropaolo2022using} is based on the fact that it is comprised solely of sampled methods without any supplementary information such as the associated path or project. This absence of information proves insufficient for conducting static analysis for obtaining context.


Next, we analyze the invocation of popular logging APIs (i.e., Log4j~\cite{log4j} and Slf4j~\cite{slf4j} at the Abstract Syntactic Tree (AST) level to extract all log statements from the original samples to complete the datasets: The extracted logging statements were marked with a line number tag (i.e., <Line Number\#>) with corresponding logging statement (i.e., <Statement>: \textit{log.info(msg)}) to indicate their position in the initial method and the complete logging statement. These served as the ground truth labels for their respective methods.

In line with works~\cite{ding2022logentext,he2018characterizing,ding2023logentext}, for each subject project, we randomly split all the methods containing logging statements into the ratio of train:test=8:2. All sampling and static analysis processes of \name will not involve any methods in the test set.

\begin{table*}[htbp]
    \scriptsize
    \vspace{-5pt}
    \centering
    \caption{Details of the studied projects.}
    \vspace{-5pt}
    \label{tab:subjects}
    \begin{tabular}{lccccc}
        \toprule
        \textbf{Project} & \textbf{Version} & \textbf{SLOC} & \textbf{\# of logging statements} & \textbf{\# of methods contain logging statements}\\
        \midrule
        ActiveMQ & 5.16.0 & 415k & 5,352 & 2,876 \\
        Ambari & 2.7.5 & 490k & 4150 & 1,689\\
        Brooklyn & 1.0.0 & 339K & 2,937 &1,374\\
        Camel & 4.0.0 & 2.12M & 9,603 & 4,460 \\
        CloudStack & 4.16.11 & 782k & 11,261 & 3,994\\
        Hadoop & 3.3.0 & 1.7M & 15,744 & 7,365 \\
        HBase & 2.4.0 & 912K & 8,677 & 3,526 \\
        Hive & 3.1.2 & 1.7M & 7,415 & 2,650\\
        Ignite & 2.8.1 & 1.1M & 4,319 & 2,335\\
        Synapse & 3.0.1 & 330k & 1,978 & 901 \\
        \bottomrule
    \end{tabular}
    \vspace{-10pt}
\end{table*}

\subsection{Baselines}
We choose LANCE~\cite{mastropaolo2022using}, the first and only one-stop logging statement approach based on T5~\cite{raffel2020exploring} and and its updated version, LANCE2.0\cite{lance2.0}, as our primary baselines, since other appraoches only focus on certain subtask (\ie logging level prediction~\cite{liu2022tell,li2021deeplv}). Code completion models are also beyond the scope of our baselines, due to their inability to infer the logging position. Given the progress in the development of Large Language Models (LLMs) and their potential use in similar development tasks, we also consider several prominent LLMs as baselines, including GPT-3.5~\cite{gpt-3.5}, Davinci~\cite{ChatGPT}, GPT-4~\cite{ChatGPT} and Llama-2-70b~\cite{touvron2023llama}. We further choose these models as the backbone models to demonstrate the generalizability of our approach. For the LLM-based baselines, we provide five fixed examples for task demonstration. Implementation details can be seen in Section.~\ref{sec:impl}.

We intentionally did not compare \name with approaches that focus on a specific aspect of \textit{what to log} (e.g., logging text generation) or current code completion models. This is because they are unable to locate the logging position in a one-stop manner and generate the corresponding logging statement. Our experimental results have demonstrated that our approach can be generalized to various backbone models, emphasizing its effectiveness as a generalized strategy instead of a specific trained model.

\subsection{Metrics}\label{sec:metric} Following the previous work\cite{mastropaolo2022using}, we evaluate the effectiveness of \name with respect to two primary dimensions: \textit{where to log} and \textit{what to log}. 

\subsubsection{Where to log}
In line with previous work~\cite{mastropaolo2022using}, we employ the metric of Position Accuracy (PA) to assess the performance of logging position prediction. We argue that the block level might be overly coarse. In this scenario, we calculate PA as 1 (indicating a successful prediction) if the distance between the predicted line number and the actual line number is less than or equal to one line and both predicted and actual line numbers must be within the same block. Otherwise, PA is calculated as 0 (indicating an unsuccessful prediction).

\subsubsection{What to log} Under the \textit{what to log} category, we evaluate \name in terms of its \textit{logging levels}, \textit{logging variables}, and \textit{logging texts} following the previous work~\cite{li2023exploring}.

(1) Logging levels.
We adopt level accuracy \textit{(L-ACC)} and Average Ordinal Distance Score \textit{(AOD)} from previous studies~\cite{li2021deeplv,liu2022tell,li2023exploring} to evaluate logging level predictions. L-ACC represents the percentage of correctly predicted log levels, while AOD calculates the distance between logging levels. since different levels are not independent of each other. For example, the \fixedwidth{error} is closer to \fixedwidth{warn} compared with \fixedwidth{trace}. The formula for \textit{AOD} is $AOD=\frac{\sum^N_{i=1} (1-Dis(a_i,s_i)/MaxDis(a_i))}{N}$, where $N$ is the number of logging statements and $MaxDis(a_i)$ refers to the maximum possible distance of the actual log level. 

(2) Logging variables.
We employ \textit{Precision}, \textit{Recall}, and \textit{F1} to evaluate the predicted set of logging variables. For each generated logging statement, we denote the variables in the model's prediction as $S_{p}$ and the variables in the actual logging statement of ground truth as $S_{g}$. We calculate the precision ($\frac{S_{p} \cap S_{gt}}{S_{p}}$), recall ($\frac{S_{p} \cap S_{g}}{S_{g}}$), and their harmonic mean (F1=$2*\frac{Precision*Recall}{Precision+Recall}$), and report these metrics. \textbf{Note that} predictions that use the same variable as the ground truth with different member function usages are considered incorrect predictions.

(3) Logging texts. 
We use \textit{BLEU}~\cite{papineni2002bleu} and \textit{ROUGE}~\cite{lin2004rouge} metrics, consistent with previous research~\cite{mastropaolo2022using,ding2022logentext}, to evaluate the quality of the generated logging texts. These metrics compute the similarity between generated and actual log messages, ranging from ~\textit{0} to ~\textit{1}. A higher score indicates better quality. Specifically, we use BLEU-K ($K=\{1, 4\}$) and ROUGE-K ($K=\{1, L\}$) to compare the overlap of K-grams.

Note that for practicality, we \textbf{only calculate} the metrics for \textit{what-to-log} when the logging position(\textit{where-to-log}), is \textbf{predicted correctly}. This is because if the logging position is not right, the purpose and meaning of the log would be incorrect, thus lacking value for further evaluation.

\subsection{Implementation}\label{sec:impl}

The static analysis part of \name has been implemented using 4,738 lines of Java code, leveraging both Soot~\cite{vallee2010soot} and Eclipse JDT Core~\cite{jdt_core} for comprehensive Java bytecode and source code analysis. The experiments of \name and all baselines were conducted on a Linux machine (Ubuntu LTS 18.04) equipped with an Intel Xeon Platinum 8255C Processor (2.50GHz), four NVIDIA A100-80GB GPUs, and 1TB of RAM.

For GPT-3.5, Davinci and GPT4, we use the public APIs provided by OpenAI\cite{ChatGPT} with \textit{gpt-3.5-turbo-0301}, \textit{text-davinci-003} and \textit{gpt-4-0314}, respectively. We run the Llama2-70b model on our machine using the Llama version \textit{Llama2-70b-chat-hf} to infer the results. By default, we set the hops of log slice and code slice to 2 and give 5 in-project examples in logging style adaption phase. For baselines, we use 5 fixed examples for task demonstration.

\vspace{-2pt}
\section{Evaluation Results}
\subsection{Research Questions}

For the evaluation, we consider the following research questions:

\begin{itemize}
  \item \textbf{RQ1}: How effective is \name compared with existing approaches?
  \item \textbf{RQ2}: What is the impact of different phases of \name?
   \item \textbf{RQ3}: How generalizable is \name for different backbone models?
  \item \textbf{RQ4}: What is the impact of different logging examples?

\end{itemize}

\subsection{RQ1: How effective is \name compared with existing approaches?}

To evaluate the effectiveness of \name in logging statement generation task, we conduct a comprehensive evaluation with comparison to other baselines on the datasets. The evaluation results are illustrated in Table.~\ref{tab:rq1-result}, where the best results for each metric are marked in \textbf{bold face}. We analyze the evaluation results from two dimensions: \textit{where-to-log} and \textit{what-to-log}. 
\begin{table*}[htbp]
    \small
    \centering
    \caption{Logging statements generation results from both ~\textit{where-to-log} and ~\textit{what-to-log} dimensions.}
        \vspace{-0.1in}
    \label{tab:rq1-result}
    \resizebox{\linewidth}{!}{%
        \begin{tabular}{l||c|cc|ccc|cccc}
            \toprule
                \textbf{Model} &
                \textbf{Posistion} &
                \multicolumn{2}{c}{\textbf{Logging Levels}} &
                \multicolumn{3}{c}{\textbf{Logging Variables}} &
                \multicolumn{4}{c}{\textbf{Logging Texts}} \\
            \cline{2-11}
                &
                \textbf{PA}    &
                \textbf{L-ACC} &
                \textbf{AOD} &
                \textbf{Precision} &
                \textbf{Recall} &
                \textbf{F1} &
                \textbf{BLEU-1} &
                \textbf{BLEU-4} &
                \textbf{ROUGE-1} &
                \textbf{ROUGE-L} \\
            \midrule
                LANCE & 0.501 & 0.574 & 0.763 & 0.657 & 0.414 & 0.508 & 0.207 & 0.110 & 0.179 & 0.175 \\
                LANCE2.0 & 0.563 & 0.601 & 0.807 & 0.632 & 0.508 & 0.563 & 0.219 & 0.113 & 0.275  & 0.266\\
                Davinci-003 & 0.307 & 0.470 & 0.714 & 0.626 & 0.544 & 0.582 & 0.267 & 0.128 & 0.288 & 0.295\\
                Llama-2-70b & 0.248 & 0.442 & 0.682 & 0.506 & 0.477 & 0.490 & 0.209 & 0.070 & 0.218 & 0.219 \\
                GPT-3.5 & 0.395 & 0.495 & 0.727 & 0.618 & 0.496 & 0.550 & 0.164 & 0.064 & 0.176 & 0.174\\
                GPT-4 & 0.518 & 0.562 & 0.779 & 0.634 & 0.611 & 0.622 & 0.285 & 0.138 & 0.317 & 0.321 \\
                \midrule
                \name & ~\textbf{0.612} & ~\textbf{0.794} & ~\textbf{0.914} & ~\textbf{0.758} & ~\textbf{0.735} & ~\textbf{0.746} & ~\textbf{0.493} & ~\textbf{0.329} & ~\textbf{0.517} & ~\textbf{0.509} \\
            \bottomrule
            \end{tabular}%
     }
\end{table*}

\noindent
\textbf{Where-to-log.} According to the evaluation results, it is clear that \name outperforms all baselines in logging position prediction. Specifically, \name outperforms the best performing baseline, LANCE2.0, by 8.7\%. Furthermore, despite having five fixed examples as task demonstrations for Large Language Models (LLMs), only GPT-4 manages to exceed the performance of the domain-specific baseline, LANCE, which is based on T5 with the considerably smaller size. The performance of the remaining LLMs on this task underscores the current limitations of these models. Notably, under the line-level metric, PA, the performance of \name further underscores the effectiveness in the task of determining \textit{where to log}. 

\noindent
\textbf{What-to-log.} We compare \name with all baselines in terms of \textit{logging levels}, \textit{logging variables}, and \textit{logging texts}. Regarding the logging levels, we observe that \name outperforms the best performing approach by 32.1\% and 13.3\% for level accuracy and AOD. When considering logging variable prediction, \name achieves a consistent improvement of 19.6\% to 20.3\% on \textit{prediction} and \textit{recall} than the best performing baseline. These results demonstrate that that \name, equipped with domain knowledge of available variables and detailed type information, significantly outperforms existing methods in predicting logging variables, along with their attributes and member functions. In terms of logging text generation, \name shows a marked improvement compared to all baselines. It achieves a BLEU-4 score of 0.386 and a ROUGE-L score of 0.599, outperforming GPT-4 by 138.4\% and 58.6\%, respectively. By adopting logging style adaptation and contextualized strategies, \name can generate logging text that aligns with the current project's logging style, both in text structure and wording. This not only improves the overall quality of the logging text but also highlights the its potential practicality for real-world software development.

\begin{tcolorbox}[boxsep=1pt,left=2pt,right=2pt,top=3pt,bottom=2pt,width=\linewidth,colback=white!95!black,boxrule=1pt, colbacktitle=white!30!black,toptitle=2pt,bottomtitle=1pt,opacitybacktitle=0.4]
\textbf{Answer to RQ1.} 
By introducing the context information in the prompt design,
\name demonstrates superior performance in both dimensions of \textit{where to log} and \textit{what to log}, significantly outperforms the best baseline. 
\end{tcolorbox}

\begin{figure}[htbp]
    \centering
    \begin{minipage}[c]{0.45\textwidth}
        \begin{subfigure}{\linewidth}
            \centering
         \includegraphics[width=\linewidth]{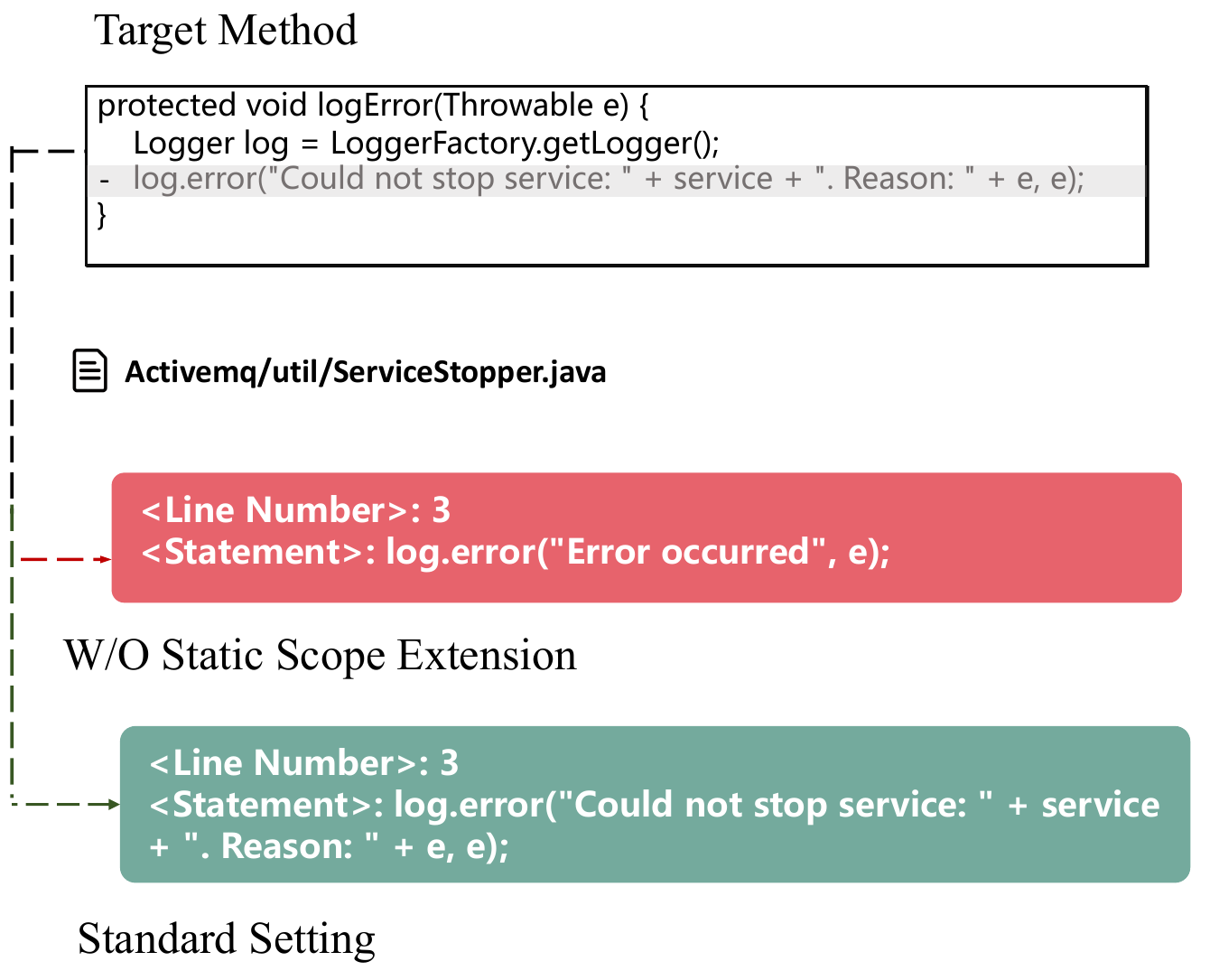}
            \caption{Removing the phase of static scope extension.}
            \label{fig:sub1}
        \end{subfigure}%
    \end{minipage}\hfill
    \begin{minipage}[c]{0.45\textwidth}
    \begin{subfigure}{\linewidth}

        \centering
        \includegraphics[width=\linewidth]{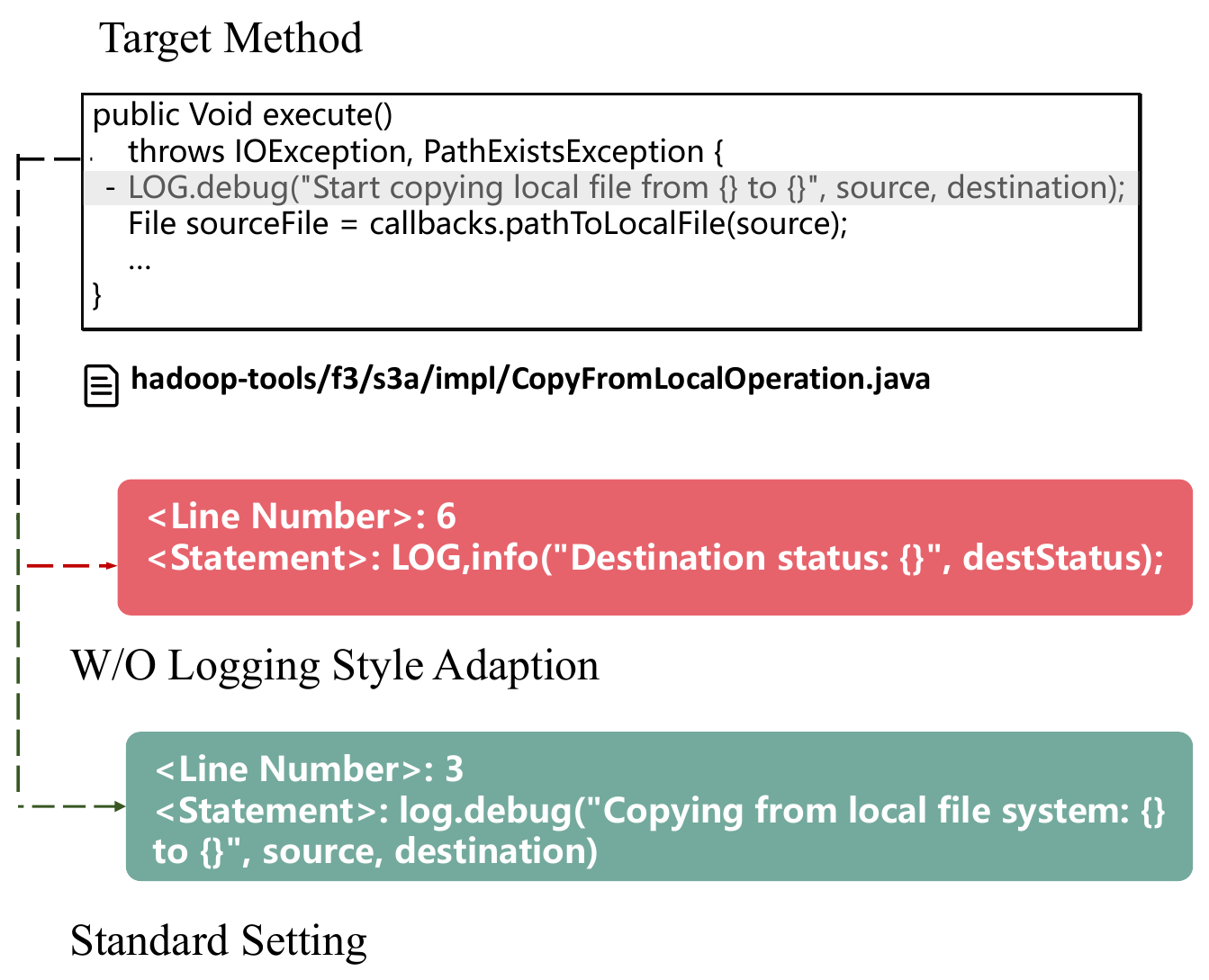}
        \caption{Removing the phase of logging style adaption.}
        \label{fig:sub2}
    \end{subfigure}
    \end{minipage}\hfill
    \vspace{-0.15in}
    \caption{Case study of the ablation study about phase static scope extension and logging static adaption.}
    \label{fig:case}
\end{figure}

\subsection{RQ2: What is the impact of different phases of \name?}

\begin{table*}[tbp]
    \small
    \centering
    \caption{Ablation Study of \name.}
        \vspace{-8pt}
    \label{tab:rq2-result}
    \resizebox{\linewidth}{!}{%
        \begin{tabular}{l||c|cc|ccc|cccc}
            \toprule
                \textbf{Ablation} &
                \textbf{Posistion} &
                \multicolumn{2}{c}{\textbf{Logging Levels}} &
                \multicolumn{3}{c}{\textbf{Logging Variables}} &
                \multicolumn{4}{c}{\textbf{Logging Texts}} \\
            \cline{2-11}
                &
                \textbf{PA}    &
                \textbf{L-ACC} &
                \textbf{AOD} &
                \textbf{Precision} &
                \textbf{Recall} &
                \textbf{F1} &
                \textbf{BLEU-1} &
                \textbf{BLEU-4} &
                \textbf{ROUGE-1} &
                \textbf{ROUGE-L} \\
            \midrule
               \name & 0.612 & 0.794 & 0.914 & 0.758 & 0.735 & 0.746 & 0.493 & 0.329 & 0.517 & 0.509\\
                w/o Loging Scope Extension & 0.579 & 0.702 & 0.858 & 0.720 & 0.711 & 0.716 & 0.430 & 0.278 & 0.468 &0.469 \\
                w/o Logging Style Adaption & 0.549 & 0.679 & 0.869& 0.752 & 0.696 &0.723 & 0.354 & 0.191 & 0.393 & 0.386\\
                w/o Logging Variable Refinement & 0.614 & 0.791 & 0.912 & 0.708 & 0.654 & 0.680 & 0.483 & 0.348 & 0.507 & 0.503\\
            \bottomrule
            \end{tabular}%
     }
     \vspace{-10pt}
\end{table*}

We conduct an ablation study to investigate the impact of different components within the framework of \name. Specifically, we design three variants of \name by removing the proposed phases \ie logging scope extension, logging style adaption and logging variable refinement in comparison with \name.

Table~\ref{tab:rq2-result} demonstrates the experimental results. 
The results indicate that without conducting the static scope extension, the overall performance of \name generally declines across all metrics. Specifically, there is a decrease of 5.4\% in position accuracy (\textit{PA}), while level accuracy (\textit{L-ACC}) and variable precision experience drops of 11.6\% and 5.0\% respectively. This decline is primarily due to the misunderstandings of the target method semantics and corresponding logging purpose, given the limited information available at the method level code. When logging style adaptation is not conducted, the performance of \name on logging text aspect experiences the most significant decrease, indicating the importance of providing logging style examples for formulating text structure and wording. Specifically, for logging text, the performance achieves the \textit{BLEU} score from 0.354 to 0.191 and the \textit{ROUGE} score from 0.393 to 0.386 for the studied projects, which are 28.2\% to 41.9\% and 24.0\% to 24.2\% lower than the standard setting, respectively. Additionally, the \textit{PA} decreases by 10.3\%, indicating that the demonstration of logging style examples can significantly enhance the performance of identifying logging positions. This also suggests that models can effectively learn logging patterns from these demonstrated examples. Furthermore, when the phase of logging variable refinement is omitted, the performance of \name for logging variables drops obviously (8.8\%, reflected by \textit{F1}), while the performance of other dimensions almost remains relatively stable (i.e., logging level). This decrease demonstrates the effectiveness of the variable refinement phase with detailed type information.

Fig.~\ref{fig:case} presents two cases (details described in Sec.~\ref{sec:example1} and Sec.~\ref{sec:example2}) to illustrate how \name can be benefited from each phase. The \textcolor{gray}{gray} line represents the original logging statements. For instance, in Fig.~\ref{fig:sub1}, without the extended method static scope information, \name failed to understand the functionality of this method. As a result, \name conservatively inferred a general error log without realizing that the current method's purpose is to report the error when the service cannot be stopped. With the help of extended static scope context (detailed in Sec.~\ref{sec:example1}), \name understand the functionality and available variables of current method, therefore can generate a more appropriate log recording the certain error. For the example shown in Fig.~\ref{fig:sub2}, \name cannot locate the logging location for such a method without understanding the logging pattern of current project. With the knowledge gaining from similar methods (detailed in Sec.~\ref{sec:example2}), \name can pinpoint the appreciate logging position (before the file operation) and generate a similar structure logging statement. The case presented in Fig.~\ref{fig: prompt} demonstrates the effectiveness of logging variable refinement. After taking the type information of the variable \textit{service}, \name can retrieve the state information and service name by calling the relevant member functions \textit{getServiceState()} and \textit{gerName()}. As a result, a more suitable logging statement is generated, and the incorrect invocation of \textit{getState()} is corrected.

\begin{tcolorbox}[boxsep=1pt,left=2pt,right=2pt,top=3pt,bottom=2pt,width=\linewidth,colback=white!95!black,boxrule=1pt, colbacktitle=white!30!black,toptitle=2pt,bottomtitle=1pt,opacitybacktitle=0.4]
\textbf{Answer to RQ2.} While evaluating individual contributions of each phase of \name, the ablation study reveals that removing any component significantly decreases the overall performance in terms of all the metrics. 
Thus, each phase individually contributes significantly to the overall effectiveness of the \name framework.
\end{tcolorbox}

\subsection{RQ3: How generalizable is \name for different backbone models?}\label{sec:ge}

\begin{table*}[tbp]
    \small
    \vspace{-5pt}
    \centering
    \caption{The performance of \name with different backbone models.}
    \vspace{-5pt}
    \label{tab:rq4-result}
    \resizebox{\linewidth}{!}{%
        \begin{tabular}{l|c|c|cc|ccc|cccc}
            \toprule
                \textbf{Model} &
                \textbf{Approach}  &
                \textbf{Posistion} &
                \multicolumn{2}{c}{\textbf{Logging Levels}} &
                \multicolumn{3}{c}{\textbf{Logging Variables}} &
                \multicolumn{4}{c}{\textbf{Logging Texts}} \\
            \cline{3-12}
                &
                &
                \textbf{PA}    &
                \textbf{L-ACC} &
                \textbf{AOD} &
                \textbf{Precision} &
                \textbf{Recall} &
                \textbf{F1} &
                \textbf{BLEU-1} &
                \textbf{BLEU-4} &
                \textbf{ROUGE-1} &
                \textbf{ROUGE-L} \\
            \midrule
               \multirow{3}{*}{LLaMa-2-70b} & Base & 0.248 & 0.442 & 0.682 & 0.506 & 0.477 & 0.490 & 0.209 & 0.070 & 0.218 & 0.219\\
               & \name & 0.282 & 0.486 & 0.743 & 0.618 & 0.467 & 0.532 & 0.283& 0.177 & 0.299 & 0.292\\
               & $\Delta$ & $\uparrow13.7\%$ & $\uparrow10.0\%$ & $\uparrow8.8\%$ & $\uparrow22.1\%$ & $\downarrow2.1\%$ & $\uparrow8.6\%$ & $\uparrow 35.4\%$ & $\uparrow152.9\%$ & $\uparrow37.2\%$ & $\uparrow33.3\%$ \\
            \midrule
            \multirow{3}{*}{GPT-3.5} & Base & 0.395 & 0.452 & 0.713 & 0.618 & 0.496 & 0.550 & 0.164 & 0.091 & 0.176 & 0.174 \\
             & \name & 0.478 & 0.559 & 0.766 & 0.712 & 0.548 & 0.619 & 0.324 & 0.213 & 0.330 & 0.329\\
             & $\Delta$ & $\uparrow21.0\%$ & $\uparrow23.7\%$ & $\uparrow7.4\%$ & $\uparrow15.2\%$ & $\uparrow10.5\%$ & $\uparrow12.5\%$ & $\uparrow97.6\%$ & $\uparrow134.1\%$ & $\uparrow87.5\%$ & $\uparrow89.1\%$\\
             \midrule
            \multirow{3}{*}{GPT-4}  & Base & 0.518 & 0.562 & 0.779 & 0.634 & 0.611 & 0.622 & 0.285 & 0.138 & 0.317 & 0.321 \\
               & \name & 0.612 & 0.794 & 0.914 & 0.758 & 0.735 & 0.746 & 
               0.493 &
               0.329 & 
               0.517 & 
               0.509 \\
            & $\Delta$ & $\uparrow18.1\%$ & $\uparrow41.3\%$ & $\uparrow17.3\%$ & $\uparrow19.6\%$ & $\uparrow20.3\%$ & $\uparrow20.3\%$ & $\uparrow73.0\%$ & $\uparrow138.4\%$ & $\uparrow63.1\%$ & $\uparrow58.6\%$\\
            \bottomrule
            \end{tabular}%
     }
     \vspace{-10pt}
\end{table*}

In this RQ, we evaluate the performance of \name by utilizing various LLMs in conjunction with our contextualized strategy. We have selected three representative and popular LLMs that are frequently used in research, specifically GPT-4, GPT-3.5, and Llama-2-70b-chat. It should be noted that by default, \name employs GPT-4 as the backbone model. Additionally, the size of the LLMs must be sufficiently large to ensure the capability of both ICL and COT~\cite{xu2023prompting,gao2023constructing}.

The experimental results are shown in Table.~\ref{tab:rq4-result}. 
We observe that our contextualized strategy can consistently enhance the performance of the utilized base models in terms of all metrics by a large margin.
On average, all models have improved by 17.6\% in determining the logging position. From the perspective of \textit{what to log}, models have improved their performance in selecting the logging level, predicting the logging variable, and generating the logging text by an average of 25\%  (reflected by \textit{L-ACC}), 13.8\% (reflected by \textit{F1}), and 60.3\% (reflected by \textit{ROUGE-L}) respectively. Meanwhile, with stronger abilities to understand our designed prompt, larger language models benefit more from the contextualized strategy associated with \name.

The results not only demonstrate the advantage of \name's design but also demonstrate the generalizability for different backbone models of our contextualized strategy. We believe that the performance of \name can be further improved with the development of code-specific large language models.

\begin{tcolorbox}[boxsep=1pt,left=2pt,right=2pt,top=3pt,bottom=2pt,width=\linewidth,colback=white!95!black,boxrule=1pt, colbacktitle=white!30!black,toptitle=2pt,bottomtitle=1pt,opacitybacktitle=0.4]
\textbf{Answer to RQ3.} 
SCLogger demonstrates the ability to consistently improve models' performance of logging statement generation, even when utilizing
relatively smaller and not code-specific language models. This demonstrates the generalizability of the proposed contextualized strategy.
\end{tcolorbox}

\subsection{RQ4: What is the impact of different logging examples?}

\begin{figure*}[htbp]
    \centering
    \includegraphics[width=\textwidth]{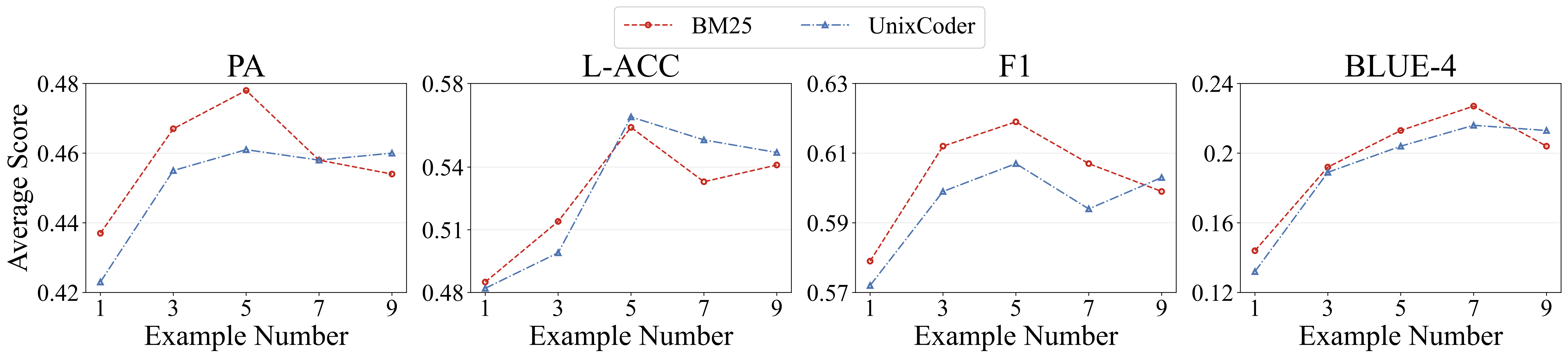}
    \vspace{-15pt}
    \caption{The selected metrics of \name with different numbers of examples and different sampling methods} 
    \label{fig: RQ4}
    \vspace{-10pt}
\end{figure*}

In this RQ, we evaluate the effects of the number of logging examples and example sampling similarity calculation approaches in the prompt design of \name. 
Following previous works~\cite{gao2023constructing,xu2023prompting,peng2023generative}, we change the number of examples from one to nine and compare two similarity calculation approaches: Unixcoder~\cite{xu2023prompting} (searching on the embedded space with Unixcoder~\cite{guo2022unixcoder}, a unified cross-modal pre-trained code large language models) and BM25. Due to the expense and limited experiment resource, the experiment is conducted using the GPT-3.5 as the backbone model.

As shown in Fig.~\ref{fig: RQ4}, we observe that the performance of \name is affected by the number of logging examples, while is less affected by the sampling approaches. The performance drops significantly when the number of examples is relative small. Moreover, the performance of \name using both sampling approaches either plateaus or starts to decrease after the example number is large than five. This proves that dissimilar logging examples and overly long prompts will result in performance loss, which stays consistent with previous works~\cite{peng2023generative, gao2023constructing} on ICL with code tasks. While comparing BM25 to UnixCoder, we observe that BM25-similarity slightly outperforms  UnixCoder in improving the performance of \name. One possible explanation is that BM25, as a text retrieval algorithm, is more capable in capturing the textual similarity instead of code semantic similarity between the logging examples and target method, thus providing more relevant examples with same syntactic structure instead of semantics.

\begin{tcolorbox}[boxsep=1pt,left=2pt,right=2pt,top=3pt,bottom=2pt,width=\linewidth,colback=white!95!black,boxrule=1pt, colbacktitle=white!30!black,toptitle=2pt,bottomtitle=1pt,opacitybacktitle=0.4]
\textbf{Answer to RQ4.} 
\name achieves the best performance with five examples, which contributes to maintaining a relatively short prompt length. The performance difference between the two sampling approaches is not significant, facilitating the use of \name in various situations.
\end{tcolorbox}
\section{Discussion}

\subsection{Practicality of \name}
\name is designed to help developers write logging statements during software development and maintenance. We discuss the practicality of \name from the following two aspects.

\noindent
\textbf{Cost reduction.} For large language models, the cost is proportional to the length of the prompt. To reduce the cost, \name only extracts and isolates the context related to logging to form the prompt, rather than taking the file-level content as input like existing programming assistants (i.e., Copilot~\cite{copilot_doc}). Moreover, for the refinement phase, \name only takes type information of chosen logging variables instead of all available variables, which also help with shorten prompt length. Our experiments show that in 84.3\% of cases, our prompt is shorter than the length of the current file of the target method, which demonstrates the relative low cost.


\noindent
\textbf{IDE integration.} \name can be easily integrated into well-established Integrated Development Environments (IDEs), such as Eclipse\cite{jdt_core}, for practical applications. In particular, Eclipse JDT\cite{jdt_core}, the built-in static analysis tool of Eclipse, has the capability to automate the majority of the static analysis procedures of \name. Compared with exisiting LLM-based code completion tools, such as Copilot~\cite{copilot_doc} or Tabnine~\cite{tabnine}, \name offers more comprehensive static features beyond method-level to improve the model's logging performance. Furthermore, experimental results (as detailed in Sec.~\ref{sec:ge}) demonstrate that \name is compatible with a variety of large language models, thereby continuously benefiting from development of LLMs.

\subsection{Threats to Validity}
\noindent
\textbf{Potential data leakage.} 
A primary concern in this work is the potential data leakage issue arising from the use of public code. Specifically, there is a possibility that the model has been trained on the test set, resulting in memorization of the results rather than conducting inference~\cite{yang2023code,rabin2023memorization, li2023exploring,huang2023not}.
To address this concern, instead of directly providing the model with the file-level contexts (which might exist in the training corpus), \name~receives a complex prompt composed of code snippets with logical reasoning relationships. This type of \textbf{data format} is unlikely to have been encountered by the model during training. Furthermore, our experimental results reveal that the model's performance in directly generating logging text is significantly below that of practical use-cases, indicating a minimal probability of direct memorization of the test set. 

\noindent
\textbf{The selection of models.} 
In this study, we employ three popular instruction-taken and practically coding-capable LLMs for experimentation, aiming to demonstrate the effectiveness of our proposed methodology. While a multitude of LLMs exist that could potentially be employed for experimentation, we have discovered that smaller parameter models fail to satisfy our requirements for understanding such complex prompt. Some models that we have experimented with either lack the capability to understand instructions or have not yet attained the level of practical application for instruction-taken coding. In future work, we plan to extend our experimentation to other emerging models, thereby evaluating the further generalizability of our method.

\noindent
\textbf{The selection of language.} One potential external concern may be that the datasets primarily rely on the Java language, which could raise questions about the generalizability of \name to other programming languages. However, Java is among the most prevalent programming languages for logging research purposes, in accordance with previous works~\cite{li2021deeplv, liu2022tell, mastropaolo2022using}. The core idea of the contextualized prompt construction and the process of static analysis can be generalized to other language with appropriate adaption.

\vspace{-4pt}
\section{Related Work}
In this section, we review the related work on empirical studies of logging practices and approaches of automatic logging. 
\vspace{-6pt}
\subsection{Studies on Logging Practices}
In order to enhance the observability and maintainability of systems, logging practices have been a subject of study~\cite{yuan2012improving, ding2015log2,chen2017characterizing,chen2021survey}, which aids developers in adopting more suitable logging strategies. \citet{fu2014developers} examined the logging practices in two large-scale online service systems involving experienced industry developers and provided six key findings about logging code categories, decision-making factors, and the feasibility of auto-logging. Furthermore, another industrial study~\cite{pecchia2015industry} revealed that logging processes are developer-dependent, highlighting the need for standardizing event logging activities across a company.
Researchers have also explored the evolution of logging statements in open software projects~\cite{chen2017characterizing, kabinna2018examining, shang2014exploratory}. These studies found that paraphrasing, inserting, and deleting logging statement operations are widespread during software development. \citet{zhao2023studying} investigated the IDs in logging statements and introduced LTID for automatic ID injection based on a log dependency graph. \citet{ding2023temporal} delved into the temporal relationships between logging and corresponding source code, leading to the detection of logging-code temporal inconsistencies through logical and semantic temporal relation rules.

Despite the comprehensive studies of logging practices, offering general and automated strategies for effective logging remains challenging. The general experience obtained from above studies is neither automatic nor consistent with the logging style of each project. To bridge the gap, this work is the first automatic one-stop logging statement generation approach with adapted logging style, benefiting further research and real-world application.

\vspace{-6pt}
\subsection{Logging Statement Automation}
Traditionally, the logging statement automaton can be divided into two steps based on stages~\cite{chen2021survey, he2021survey}: the selection of logging locations and the generation of logging statements, which we summarize as \textit{where-to-log} and 
\textit{what-to-log}. 

To solve the problem of \textit{where to log}, researchers have tried many approaches~\cite{li2020shall,zhao2017log20,zhu2015learning,yao2018log4perf,jia2018smartlog} to find the appropriate logging location in the source code. Prior studies~\cite{lal2016logoptplus,yuan2012conservative} have tackled the log placement problem within specific code constructs such as \textit{catch} and \textit{if} statements. However, such logging placement can lead to an excess of logging statements, bringing additional system overhead. Log20~\cite{zhao2017log20} was proposed to identify an almost optimal placement of logging statements, guided by information theory and under the constraints of performance overhead. It determines the logging position by evaluating the effectiveness of each logging statement in distinguishing execution paths. With the development of machine learning, data-driven approaches~\cite{zhu2015learning,li2020shall} have brought about more possibilities. LogAdvisor~\cite{zhu2015learning} will suggest logging positions by learning structural features, textual features, and syntactic features from systems. By introducing deep learning to learn the features from source code, \citet{li2020shall} has elevated performance to a new level.

For \textit{what to log}, the process of generating logging statements is generally divided into three subtasks: logging level prediction~\cite{li2021deeplv, liu2022tell, li2017log}, logging variables selection~\cite{liu2019variables, dai2022reval, yuan2012improving}, and logging text generation~\cite{mastropaolo2022using, ding2022logentext}. Ordinal-based neural networks~\cite{li2021deeplv} and graph neural networks~\cite{liu2022tell} have been utilized to learn syntactic code features and semantic text features to recommend for logging level. LogEnhancer~\cite{yuan2012improving}, from a programming analysis viewpoint, aspires to alleviate the complexity of failure diagnosis by incorporating causally-related variables into a logging statement. Meanwhile, LoGenText~\cite{ding2022logentext} and LoGenText-Plus~\cite{ding2023logentext} translates the related source code into short textual descriptions and then generate the logging text using neural machine translation models.

The most recent approach LANCE~\cite{mastropaolo2022using}, provides a one-stop logging statements solution of deciding logging points and logging statements for method level java code. Nevertheless, owing to its end-to-end design and limited method-level context, the approach suffers from insufficient context. As a result, it fails to satisfy the practical needs of real-world development scenarios. To address this, our approach for context-aware logging statement generation simultaneously addresses the issues of \textit{where-to-log} and \textit{what-to-log}, providing a practical solution for logging statement automation.

\vspace{-4pt}
\section{Conclusion}
In this paper, we propose \name, the first contextualized logging statement generation approach with static contexts.
\name incorporates static domain knowledge into language models via a context-aware prompt structure and further self-refinement. Experimental results show that \name outperforms all baselines and can be generalized various LLMs. We believe
that \name would benefit both developers and researchers in the field of logging statement generation.

\vspace{-4pt}

\section*{Acknowledgment}

The work described in this paper was supported by the Research Grants Council of the Hong Kong Special Administrative Region, China (No. CUHK 14206921 of the General Research Fund). We thank all reviewers for their valuable comments.

\bibliographystyle{ACM-Reference-Format}
\bibliography{sample-base}

\end{document}